\documentclass[aps,pra,final,10pt,twocolumn,superscriptaddress]{revtex4-1}
\usepackage{amsmath,amssymb,bm}
\usepackage{graphicx,color}
\usepackage[squaren]{SIunits}
\usepackage[english]{babel}
\usepackage{amstext}
\usepackage{amsthm}
\usepackage{latexsym}
\usepackage{array}
\usepackage{color}
\usepackage{float}
\usepackage{microtype}
 
\usepackage{multirow}
\usepackage{dcolumn}
\definecolor{url}{RGB}{0,20,160}
\usepackage[colorlinks=true,linkcolor=blue,citecolor=blue,urlcolor=url]{hyperref}
\usepackage[usenames,dvipsnames,svgnaes,table]{xcolor}
\usepackage{graphicx, type1cm, lettrine}
\usepackage[utf8]{inputenc}
\usepackage{Typocaps}
\usepackage[T1]{fontenc}
\usepackage{palatino}

\usepackage{natbib}
\makeatletter
\def\frutiger{cmss10 }
\def\frutigerbold{cmssbx10 }
=\frutigerbold at 14pt
=\frutiger at 8pt
=\frutigerbold at 12pt
=\frutigerbold at10pt
=\frutigerbold at 8pt
=\frutiger at 8pt
=\frutigerbold at 31pt
\def\@caption@tabnum@sep{\figtextfont{{ }{\bf\textbar}{ }}}%
\def\fnum@table{{\bf\tablename~\thetable}}
\renewenvironment{table}{\@float{table}\def\textbf##1{{\fignumfont ##1}}\def\bf{\fignumfont}}{\end@float}
\def\@caption@fignum@sep{\figtextfont{{ }{\bf\textbar}{ }}}%
\def\fnum@figure{{\bf\figurename~\thefigure}}
\renewenvironment{figure}{\@float{figure}\def\textbf##1{{\fignumfont ##1}}\def\bf{\fignumfont}}{\end@float}
\def\@startsection#1#2#3#4#5#6{%
\if@noskipsec\leavevmode\fi
\par\@tempskipa #4\relax
\@afterindenttrue
\ifdim\@tempskipa <\z@
\@tempskipa -\@tempskipa \@afterindentfalse
\fi\if@nobreak\everypar{}%
\else\addpenalty\@secpenalty\addvspace\@tempskipa\fi
\@ifstar{\@ssect{#3}{#4}{#5}{#6}}{\@dblarg{\@sect{#1}{#2}{#3}{#4}{#5}{#6}}}}
\def\@sect#1#2#3#4#5#6[#7]#8{%
\ifnum #2>0
\let\@svsec\@empty
\else\refstepcounter{#1}\protected@edef\@svsec{\@seccntformat{#1}\relax}\fi
\@tempskipa #5\relax
\ifdim\@tempskipa>\z@
\begingroup#6{\@hangfrom{\hskip #3\relax\@svsec}%
\interlinepenalty \@M #8\@@par}\endgroup
\csname #1mark\endcsname{#7}%
\addcontentsline{toc}{#1}{%
\ifnum #2>\c@secnumdepth\else
\protect\numberline{\csname the#1\endcsname}\fi #7}%
\else\def\@svsechd{#6{\hskip #3\relax
\@svsec #8\ifnum#2=2.\fi}%
\csname #1mark\endcsname{#7}%
\addcontentsline{toc}{#1}{%
\ifnum #2>\c@secnumdepth \else
\protect\numberline{\csname the#1\endcsname}\fi #7}}%
\fi\@xsect{#5}}

\renewcommand\section{\@startsection {section}{1}{\z@}%
{-10pt \@plus -1ex \@minus -.2ex}{.5ex }{\normalfont\Large\bfseries\sectionfont}}
\renewcommand\subsection{\@startsection{subsection}{2}{\z@}%
{10pt\@plus 1ex \@minus .2ex}{-0.5ex \@plus .2ex}{\normalfont\large\bfseries\subsectionfont}}
\def\frontmatter@title@format{\titlefont\centering}%
\def\frontmatter@title@below{\addvspace{-5pt}}%

\renewcommand\NAT@biblabelnum[1]{#1.}
\renewcommand\NAT@citesuper[3]{\ifNAT@swa
\unskip\hspace{1\p@}\textsuperscript{(#1)}%
   \if\relax#3\relax\else\ (#3)\fi\else (#1)\fi\endgroup}

\newcommand*\bib@heading{%
  \section{\refname}
  \fontsize{8}{10}\selectfont
}
\newcommand*\@openbib@code{%
      \advance\leftmargin\bibindent
      \itemindent -\bibindent
      \listparindent \itemindent
      \parsep \z@
}%
\newdimen\bibindent
\usepackage[usenames,dvipsnames,svgnaes,table]{xcolor}
\definecolor{col1}{rgb}{0.0, 0.30, 1.0}
\definecolor{col2}{rgb}{0.9, 0.0, 0.30}

\makeatletter
\begin{document}
\title{Designing and discovering a new family of semiconducting quaternary Heusler
\\[0.25\baselineskip] compounds based on the 18-electron rule}
\author{Jiangang He}
\affiliation{Department of Materials Science and Engineering, Northwestern University, Evanston, Illinois 60208, United States}
\author{S. Shahab Naghavi}
\affiliation{Department of Physical and Computational Chemistry, Shahid Beheshti University, G.C., Evin, 1983969411 Tehran, Iran}
\affiliation{Department of Materials Science and Engineering, Northwestern University, Evanston, Illinois 60208, United States}
\author{Vinay I. Hegde}
\affiliation{Department of Materials Science and Engineering, Northwestern University, Evanston, Illinois 60208, United States}
\author{Maximilian Amsler}
\altaffiliation[Current address: ]{Laboratory of Atomic and Solid State Physics, Cornell University, Ithaca, New York 14853, USA}
\author{Chris Wolverton}
\email{c-wolverton@northwestern.edu}
\affiliation{Department of Materials Science and Engineering, Northwestern University, Evanston, Illinois 60208, United States}


\date{\today}

\begin{abstract}
\noindent \textbf{Intermetallic compounds with sizable band gaps are attractive
for their unusual properties but rare. Here, we present a new family of stable
semiconducting quaternary Heusler compounds, designed and discovered by means of
high-throughput \textit{ab initio} calculations based on the 18-electron rule.
The 99 new semiconductors reported here adopt the ordered quaternary Heusler
structure with the prototype of LiMgSnPd (F$\bar{\mathbf{4}}$3m, No.\,216) and
contain 18 valence electrons per formula unit. They are realized by filling the
void in the half Heusler structure with a small and electropositive atom, i.e.,
lithium. These new stable quaternary Heusler semiconductors possess a range of
band gaps from 0.3 to 2.5\,eV, and exhibit some unusual properties different from conventional
semiconductors, such as strong optical absorption, giant dielectric screening,
and high Seebeck coefficient, which suggest these semiconductors
have potential applications as photovoltaic and thermoelectric materials.
While this study opens up avenues for further exploration of this novel class of semiconducting quaternary Heuslers, the design strategy used herein is broadly applicable across a potentially wide array of chemistries to discover new stable materials.}
\end{abstract}
\maketitle

\begin{figure}
\includegraphics[clip,width=1.0\linewidth]{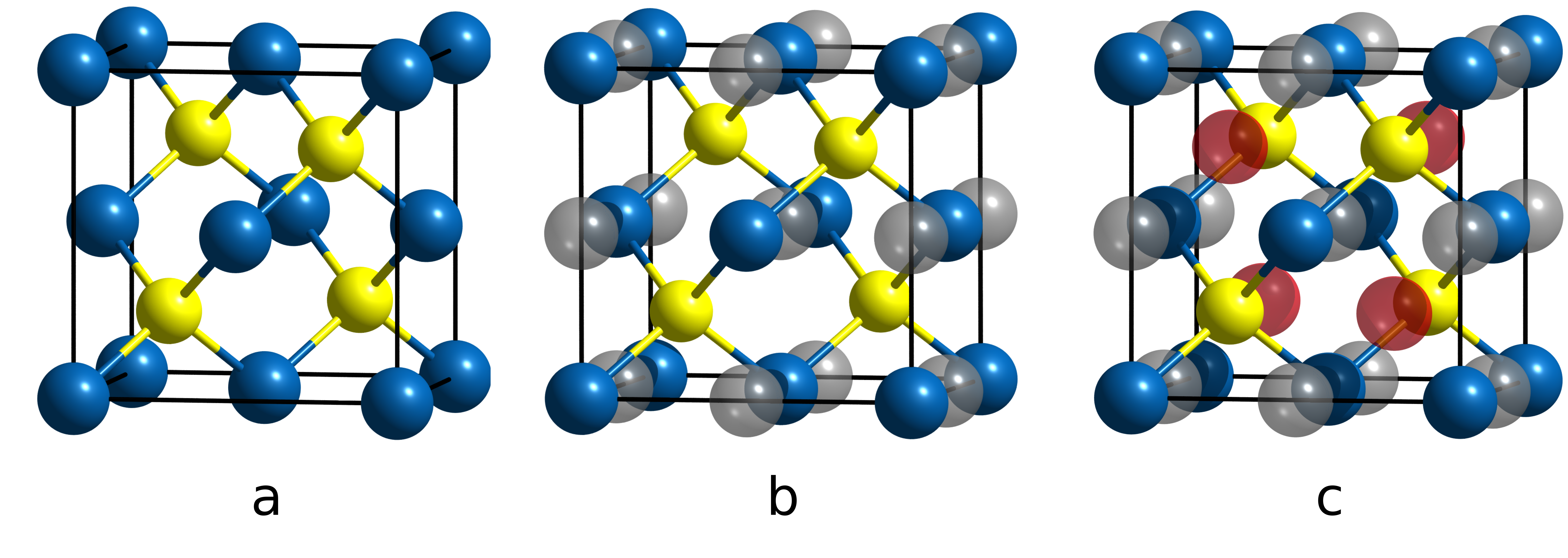}
\caption {\textbf{Comparison between zinc blende and Heusler structures.}
\textbf{a.} $XZ$ forms a zinc blende lattice. \textbf{b.} The half Heusler $XYZ$
with $Y$ occupying the octahedral void in the zinc blende lattice $XZ$.
\textbf{c.} The quaternary Heusler $XX^{\prime}YZ$ with $X^{\prime}$ occupying
the tetrahedral void of the half Heusler $XYZ$. $X$, $X^{\prime}$, $Y$, $Z$
atoms are yellow, red, gray, and blue, respectively.}
\label{FIG:ZnS}
\end{figure}

\lettrine[lines=3, findent=3pt, nindent=0pt]{\color{BurntOrange}H}{}eusler
compounds are one of the largest families of ternary intermetallic compounds and
have been intensely studied due to their simple crystal structure and broad
range of applications~\cite{heusler,dshemuchadse2015more,Graf2011}. The
electronic structure and magnetic properties of Heusler compounds can be very
well described by valence electron counting rules, i.e., the Slater-Pauling rule
and 18-electron
rule~\cite{PhysRevB.66.174429,Graf2011,Kandpal2006,FelserBook2016}, enabling
rational material design. Ternary Heusler compounds exist in three well-known
varieties: full Heusler $X_2YZ$ (FH); inverse Heusler $XY_2Z$ (IH); and half
Heusler $XYZ$ (HH). In contrast to the intensively studied Ternary
Heusler~\cite{Graf2011,PhysRevB.66.174429,PhysRevB.66.134428,Gautier2015,PhysRevX.4.011019,sanvito2017accelerated,PhysRevMaterials.1.034404}, quaternary Heusler $XX'YZ$ compounds (QH)
have not been as well explored yet, presumably due to the complexity of the
quaternary phase space. As illustrated in Fig.~\ref{FIG:ZnS}, all these Heusler
structures can be interpreted in terms of their relationship to the zinc blende
$XZ$ structure (space group F$\bar{4}3m$, No.~216; Wyckoff positions: $X$
($4a$): (0, 0, 0) and $Z$ (4c): (0.25, 0.25, 0.25); see Fig\,\ref{FIG:ZnS}a),
where both the octahedral $4b$ (0.50, 0.50, 0.50) and tetrahedral $4d$ (0.75,
0.75, 0.75) sites are unoccupied. HH $XYZ$ is derived by filling the octahedral
site (Wyckoff site $4b$) in the zinc blende $XZ$ structure with a $Y$ atom
(Fig\,\ref{FIG:ZnS}b). Further filling the empty tetrahedral sites with another
atom $X^{\prime}$ results in the QH $XX^{\prime}YZ$ structure (Fig\,\ref{FIG:ZnS}c). FH $X_{2}YX$ is generated when $X^{\prime}$ and $X$ are the same element, and subsequently its crystal symmetry increases to Fm$\bar{3}m$, whereas IH $XY_2Z$ is formed by substituting the $X^{\prime}$ atom with $Y$ in the QH structure. \\
Most Heusler intermetallics have a metallic band structure and typically, only
compounds with specific numbers of valence electrons can be
semiconductors~\cite{PhysRevB.66.174429,Graf2011,PhysRevLett.117.046602}. For
instance, FH compounds containing 24 valence electrons per formula unit (f.u.)\
or HH compounds with 18 valence electrons per f.u.\ are likely to be
semiconductors~\cite{Graf2011}. The band gap of these Heusler compounds is
mainly determined by the magnitude of $Y$-site $d$ orbital splitting in the
crystal field and the hybridization between $X$-site $d$- and $Z$-site
$p$ orbitals~\cite{Graf2011,Kandpal2006,Yan2015,FelserBook2016,Chadov2010,Zeier2016}.
Many HH and FH semiconductors have been discovered and intensively studied for
various applications such as thermoelectrics, transparent conductors,
topological insulators, and buffer layers of solar
cells~\cite{Zhu2015,Zeier2016,Yan2015,Kieven2010,Meinert2014,PhysRevLett.114.136601,PhysRevLett.117.046602}.
Due to the plethora of exotic physical properties, the Heusler family has become
a playground for the theoretical investigation of novel compounds. Recently, for
instance, Gautier et\,al.\ systematically explored missing 18-electron $XYZ$
ternary compounds by means of high-throughput (HT) \emph{ab initio} virtual
screening and discovered 20 new stable HH semiconductors with band gaps ranging
from 1.5\,eV to 3.0\,eV from a dataset of 400 compounds~\cite{Gautier2015,Yan2015}.
Many of these predicted compounds were subsequently synthesized
experimentally---a powerful demonstration and validation of the HT computational
approach.

Compared with the ternary Heusler (FH, HH, and IH) compounds, the quaternary
phase space of the QH compounds affords a much higher freedom in tuning and
designing electronic and magnetic structures for optical, electronics, and
magnetic applications~\cite{PhysRevB.91.104408}. However, the number of the
potential $XX^{\prime}YZ$ compounds is 3,265,290 (C$_4^{73}\times 3$; 73 is the
number of metallic elements~\cite{Kirklin2016}, and there are 3 possible
structures for each $XX^{\prime}YZ$ composition), which makes an exhaustive
experimental search impractical. HT computation based on density functional
theory (DFT) can screen this large phase space to identify promising candidates
that are thermodynamically and dynamically stable, and in addition predict their
ground state crystal structure and interesting properties, thereby guiding and
accelerating experimental efforts aimed at their successful
synthesis~\cite{curtarolo2013high,Gautier2015,Yan2015,sanvito2017accelerated}.

To search for new QH semiconductors efficiently, in this paper, we constrain our
HT DFT screening to the Li-containing quaternary phase space (Li-$X$-$Y$-$Z$),
including 53 possible elements in the $X$, $Y$, $Z$ sites (elements with
partially occupied $f$ orbitals are excluded to avoid problems with convergence,
see Supplementary for the full list), and considering compounds
with 18 valence electron per f.u. The small atomic size of
Li is very suitable for filling the void in a HH structure, \textit{and} the
electron donated by the electropositive Li can fully fill the valence band of the 17-electron HH
$XYZ$. Hence, the corresponding 18-electron QH system obtained in this way
is stabilized by fully filling
bonding states, and accompanied by a band gap opening between the valence
bonding states and the conduction anti-bonding states. Using HT DFT calculations
within the framework of the Open Quantum Material Database
(OQMD)~\cite{OQMD,kirklin2015open}, and first-principles phonon calculations of
1320 candidate compounds, we find 99 Li-containing QH intermetallic
semiconductors and 5 metals that are thermodynamically and dynamically stable.
Based on our calculations, these compounds possess a wide range of band gaps
(from 0.31 to 2.25\,eV) with some candidates possessing exciting properties such
as strong visible light absorption, giant high-frequency (electron) dielectric
function, and high power factor. We hereby predict several promising candidates
as energy harvesting materials for photovoltaic and thermoelectric applications.

\begin{figure*}[htp!]
\centering
\includegraphics[width=1.0\linewidth]{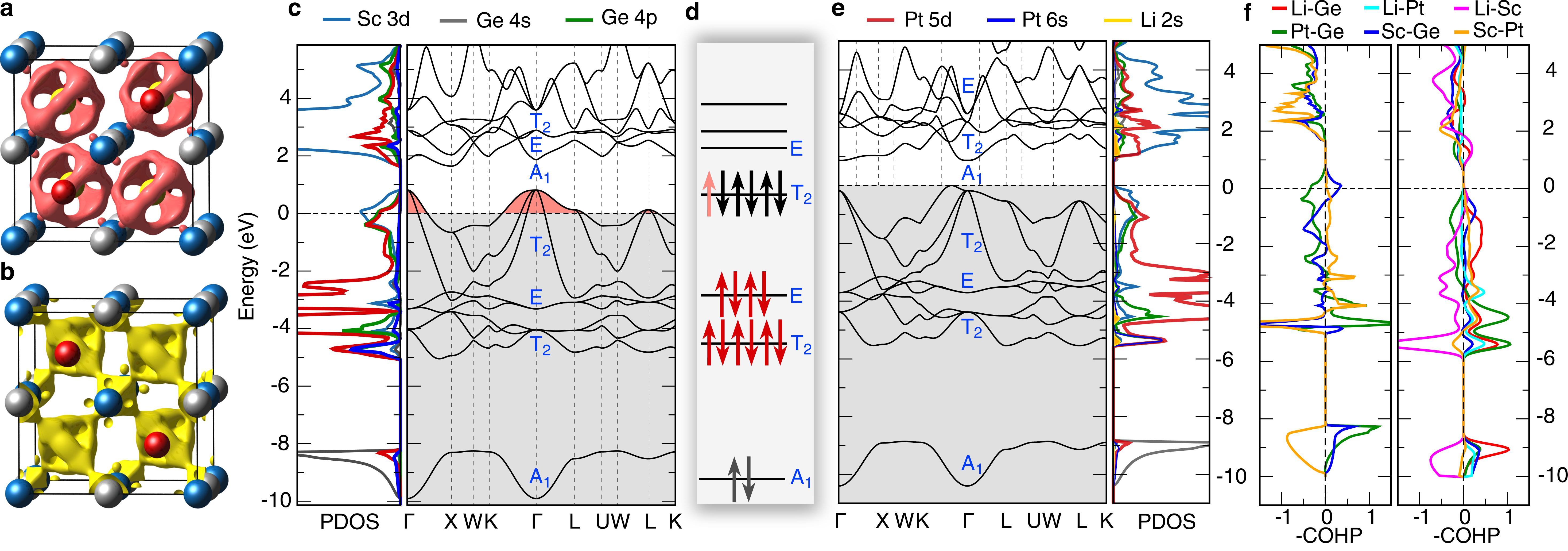}
\caption{{\bf \sffamily Electronic structure of a 17-electron half Heusler
(PtScGe) vs.\ the corresponding 18-electron quaternary Heusler (PtLiScGe).}
\textbf{\sffamily a.}  Charge density difference between PtLiScGe and PtScGe.
\textbf{\sffamily b.} Electron localization function (ELF) of PtLiScGe.
\textbf{\sffamily c.} band structure and partial density of states (PDOS) of
PtScGe. \textbf{\sffamily e.} band structure and PDOS of PtLiScGe.
\textbf{\sffamily f.} crystal orbital Hamilton populations (-COHP) of PtScGe.
\textbf{\sffamily g.} -COHP of PtLiScGe.}
\label{FIG:BAND}
\end{figure*}

\section{Design rules for semiconducting quaternary Heusler compounds}
The emergence of a finite band gap in conventional semiconductors/insulators,
such as Si, GaN, ZnS, and NaCl can be understood by the well-known octet
rule~\cite{lewis1916atom,langmuir1919arrangement}, where the fully occupied $s$
and $p$ orbitals form a $s^2p^6$ closed shell. Owing to the close connection
between the zinc blende (ZnS) and HH structure as mentioned before, the octet rule
is also applicable to main-group-element-based HH, such as LiMgAs and LiAlSi.
Since transition metals have more low-lying energy states and $d$ electrons, an
18-electron closed shell ($s^2p^6d^{10}$) is required to open the band gap for a
HH compound with one or two transition metal atoms; this is known as the 
18-electron rule~\cite{mingos2004complementary,Gautier2015}. The valence and
conduction bands of HH, taking PtScGe as an example, are a consequence of the
hybridization between Sc ($Y$-site) 3$d$ orbitals and Ge ($Z$-site) 4$p$
orbitals as depicted in Fig.~\ref{FIG:BAND}c. Owing to the presence of a
tetrahedral crystal field and the absence of inversion symmetry (T$_{\rm{d}}$
point group), the $d$ orbitals of the transition metal atoms Pt and Sc split
into T$_2$ and E orbitals at the $\Gamma$ point of the Brillouin zone, resulting
in 10 orbitals, T$_2$ (denoted as T$_2^{1}$), E (denoted as E$^1$), T$_2$ (noted as T$_2^{2}$),
and E (noted as E$^2$) from low to high energy. The band gap
appears between the T$_2^2$ and E$^2$ orbitals that are mainly
from Sc. The A$_{1}$ band below the T$_2^1$ stems from the 4$s$ orbital of
the main group element Ge. Therefore, when the total valence electron count of
the HH system is 18, the orbitals A$_1$, T$_2^1$, E$^1$, and T$_2^2$ are
fully occupied and the orbitals higher than the T$_2^2$ are
completely empty. As a consequence, the HH compound shows a semiconducting band
structure. The size of the band gap mainly depends on the splitting between the
T$_2^2$ and E$^2$ bands and the strength of hybridization between these
orbitals. As expected, the 17-electron HH PtScGe is metallic with the Fermi
level crossing the T$_2^2$ band as demonstrated in Fig.~\ref{FIG:BAND}c. A
semiconducting state is achieved by filling the void in the HH $XYZ$ compound
with a fourth atom $X'$ that possesses a suitable ionic radius, to minimize the
strain, and a strong electropositivity, to donate its electron to other atoms
thereby raising the Fermi level to the top of the valence band. Lithium is a
good candidate to satisfy the above two conditions at the same time. As shown in
Fig.~\ref{FIG:BAND}e, the Li filled HH PtScGe, namely PtLiScGe, indeed has a
band structure very similar to the corresponding 17-electron HH counterpart: the
valence and conduction bands are mainly from Sc ($Y$-site) 3$d$ and Ge
($Z$-site) 4$p$ orbitals, respectively, whereas the Pt ($Y$-site) 5$d$ states are
deeply buried below the Ge 4$p$ states. One 2$s$ electron from Li fills the
empty portion of the valence band of PtScGe, resulting in a semiconductor with fully occupied
valence bands and completely empty conduction bands.
Owing to the high electropositivity, the Li atom donates its 2$s$ electron
to the more electronegative atoms, i.e., Ge and Pt. The delocalization of Li
2$s$ can be clearly seen from the partial density of states (PDOS) shown in
Fig.~\ref{FIG:BAND}e.

Figs.~\ref{FIG:BAND}f--g show the projected crystal orbital Hamilton populations
(pCOHP) of PtScGe and PtLiScGe, which reflects the bonding interactions between
various pairs of species in each compound. It is clear that Li--Ge and Li--Pt
pairs have a strong bonding interaction below the Fermi level due to the charge
transfer from Li to Ge and Pt, consistent with the large electronegativity
difference between them. The bonding interactions between Pt and Ge do not
change between the HH and QH except a downward shift with respect to Fermi
level, whereas the anti-bonding interactions between Sc--Pt and Sc--Ge pairs are
significantly reduced after inserting the Li atom. The main anti-bonding
interaction in PtLiScGe is the Li--Sc pair, which are both strong electron
donors, consistent with the fact that there are no stable Li--Sc binary
compounds known.

\begin{figure*}
\centering
\includegraphics[width=1.0\linewidth]{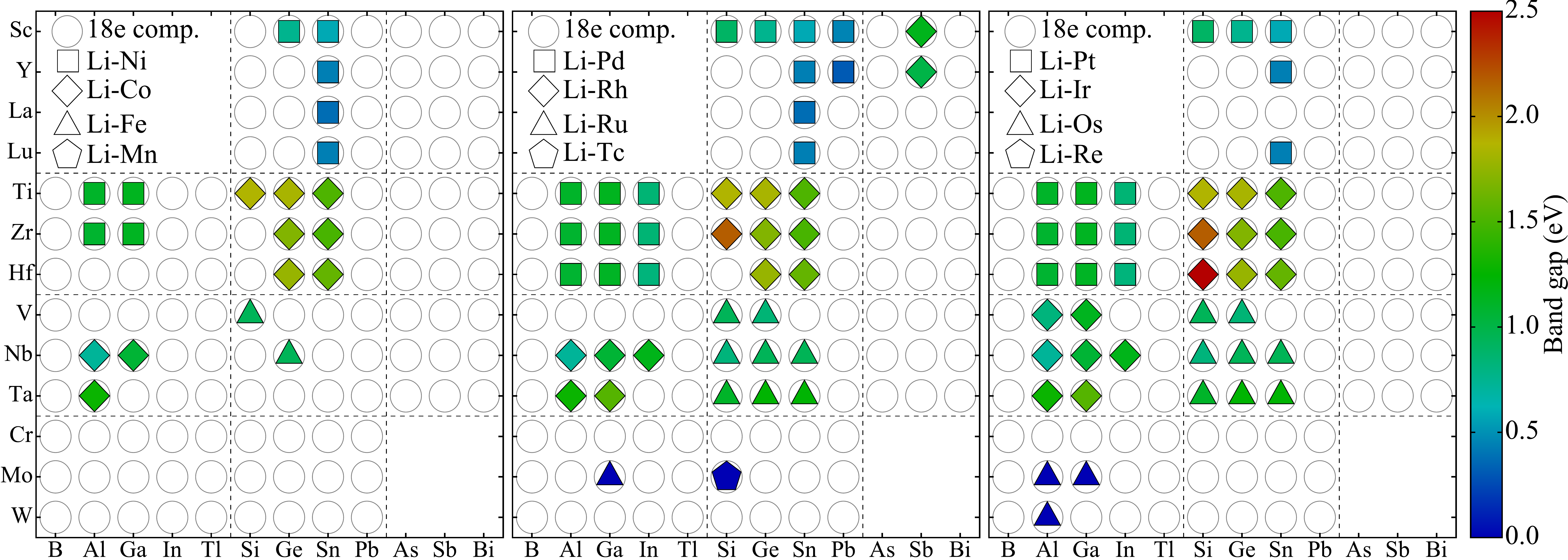}
\caption{\textbf{\textsf{Heat map of band gap (E$_{\mathbf{g}}$) for stable
Li-containing 18-electron QH compounds considered in this work.}} $X$Li$YZ$
compound is defined by the Li--$X$ and $Y$--$Z$ sublattices. Square, diamond,
triangle, and pentagon shapes correspond to different Li--$X$ pairs with $X=$
group 10, 9, 8, and 7 elements, respectively. Each circle represents an
18-electron compounds.}
\label{FIG:HTP}
\end{figure*}

\section{Thermodynamic stability and lattice dynamics}\label{sec:stability}

Thermodynamic stability is the most compelling and challenging criterion that
has to be considered in predicting new compounds. In this paper, the
thermodynamic stability is critically evaluated with respect to: (a) all the
stable phases within the Li-$X$-$Y$-$Z$ phase space, (b) chemical disorder
within one rock-salt sublattice and between two rock-salt sublattices (see Supplementary
for details), and (c) competing quaternary crystal structures for a given
composition $X$Li$YZ$. Condition (a) was evaluated by constructing the convex
hull of formation energies of all the competing compounds included in the
OQMD~\cite{OQMD,kirklin2015open}, which includes energetics of over 600,000
compounds, either reported in the Inorganic Crystal Structure Database (ICSD) or
constructed from common binary and ternary prototype
structures~\cite{Kirklin2016}. We find 99 $X$Li$YZ$ compounds to lie on the
respective formation energy convex hull, implying that they are
thermodynamically stable at zero Kelvin.

Since site occupancy disorder is commonly observed in Heusler compounds, the
total energies of the 0 K stable QH phase were further compared with the
disordered one simulated using special quasirandom structures (SQS) (see Supplementary for
details). We find that ordered QH compounds always have much lower total
energies than the corresponding disordered compounds (by $\sim$100 meV/atom; for
scale, the ideal entropy of mixing $\Delta$S$_{\rm mix}$ = log$_{\rm e}$2
k$_{\rm B}$, which is $\sim$18\,meV at 300\,K. See Supplementary Table~\textcolor{blue}{S1}
for details), indicating that most of the $X$Li$YZ$ compounds are indeed
ordered.

Nevertheless, the QH structure is not necessarily the ground state crystal
structure of the $X$Li$YZ$ composition as other structures might be
lower in energy. To check whether other lower-energy crystal structures exist at
a given $X$Li$YZ$ composition, we performed global structure searches based on
(a) Minima Hopping Method (MHM) as implemented in the Minhocao
package~\cite{mhm1,mhm2}, and (b) particle-swarm optimization technique as
implemented in the CALYPSO package~\cite{PhysRevB.82.094116,wang2012calypso}.
Since crystal structure prediction searches are computationally very expensive,
around 10\% QH compounds were randomly selected for these more refined searches.
As tabulated in Supplementary Table~\textcolor{blue}{S1}, no other lower energy crystal
structure was found for any of the selected QH compounds.

Furthermore, we examined the lattice dynamical stability of all the 99 QH
compounds by performing phonon and elastic constant calculations (see the
\textcolor{blue}{Methods} section for details). The results show that all the newly
discovered $X$Li$YZ$ QH compounds are lattice dynamically stable. Therefore, we
can safely claim that all $X$Li$YZ$ compounds are thermodynamically and lattice
dynamically stable in the ordered QH structure.

The distribution of the 99 Li-containing QH compounds in terms of chemistry is
shown in Fig.~\ref{FIG:HTP}. The site preference for different elements in
these stable QH are uniform: Li and a late transition metal (Li--$X$) form one
rock-salt sublattice, and an early transition metal and a main group element
($Y$-$Z$) form the other rock-salt sublattice. This elemental preference for an
atomic site is very similar to a HH compound~\cite{Gautier2015}, where a vacancy
and a late transition metal form one rock-salt sublattice. These findings are
also consistent with our material design strategy, namely, the vacant site of a
HH is filled with the Li atom which, on account of its electropositivity,
readily bonds with the more electronegative late transition metal. Finally, the
most stable QH compounds are mainly from the following groups, $X$: group 9 and
10; $Y$: group 3, 4, and 5; $Z$: group 13 and 14. The complete distribution of
formation energy and convex hull distances for all the QH compounds studied in 
this paper are provided in the Supplementary Figs.~\textcolor{blue}{S1} and ~\textcolor{blue}{S2}.

\section{Electronic structure}
After screening for thermodynamic and lattice dynamic stabilities, we calculated
the band gaps of all the stable QH compounds using the screened hybrid
functional HSE06~\cite{heyd2003hybrid}, since semilocal exchange-correlation
functionals such as Perdew-Burke-Ernzerhof (PBE)~\cite{PBE} tend to
underestimate band gaps. As shown in Fig.~\ref{FIG:HTP} all compounds are
semiconductors with band gaps ranging from 0.3 to 2.5\,eV except for Mo and W
based QH (RuLiMoGa, OsLiMoAl, OsLiWAl, OsLiMoAl, and TcLiMoGa), which are
metals. Among the semiconducting QH compounds, HfLiSiIr and PdLiYPb have the
largest (2.5\,eV) and smallest band gaps (0.3\,eV), respectively. We see from
Fig.~\ref{FIG:HTP} that the QH compounds containing elements from group 4 and 14
generally have larger band gaps than other compounds, and the metallic compounds
are the ones containing a group 6 element. The magnitude of the band gap has a strong
correlation with the electronegativity of $X$, $Y$, and $Z$-site elements, and
the $d$ electron occupation of $X$ and $Y$ elements. All semiconductors have
qualitatively similar band structure to PtLiScGe (Fig.~\ref{FIG:BAND}e). Although
Mo and W based QH are metals, the fundamental characteristics of their band
structure are similar to that of PtLiScGe as well. The metallic nature
is mainly a consequence of the weak hybridization between the $X$ and $Y$
elements due to a small difference in electronegativity. The electronic structure of
our predicted QH compounds is quite different from conventional semiconductors,
such as GaAs and ZnS, in that both the valence band and conduction bands are
mainly from $d$ orbitals. For all QH studied in this paper, the density of
states around the Fermi level, which stems from the spatially localized $d$
orbitals, is very high. It is therefore expected that these QH compounds have
interesting applications as solar cells due to strong light absorption and
thermoelectric materials because of the high thermopower, which will be
discussed in the following two sections.

\begin{figure}
\centering
\includegraphics[width=0.8\linewidth]{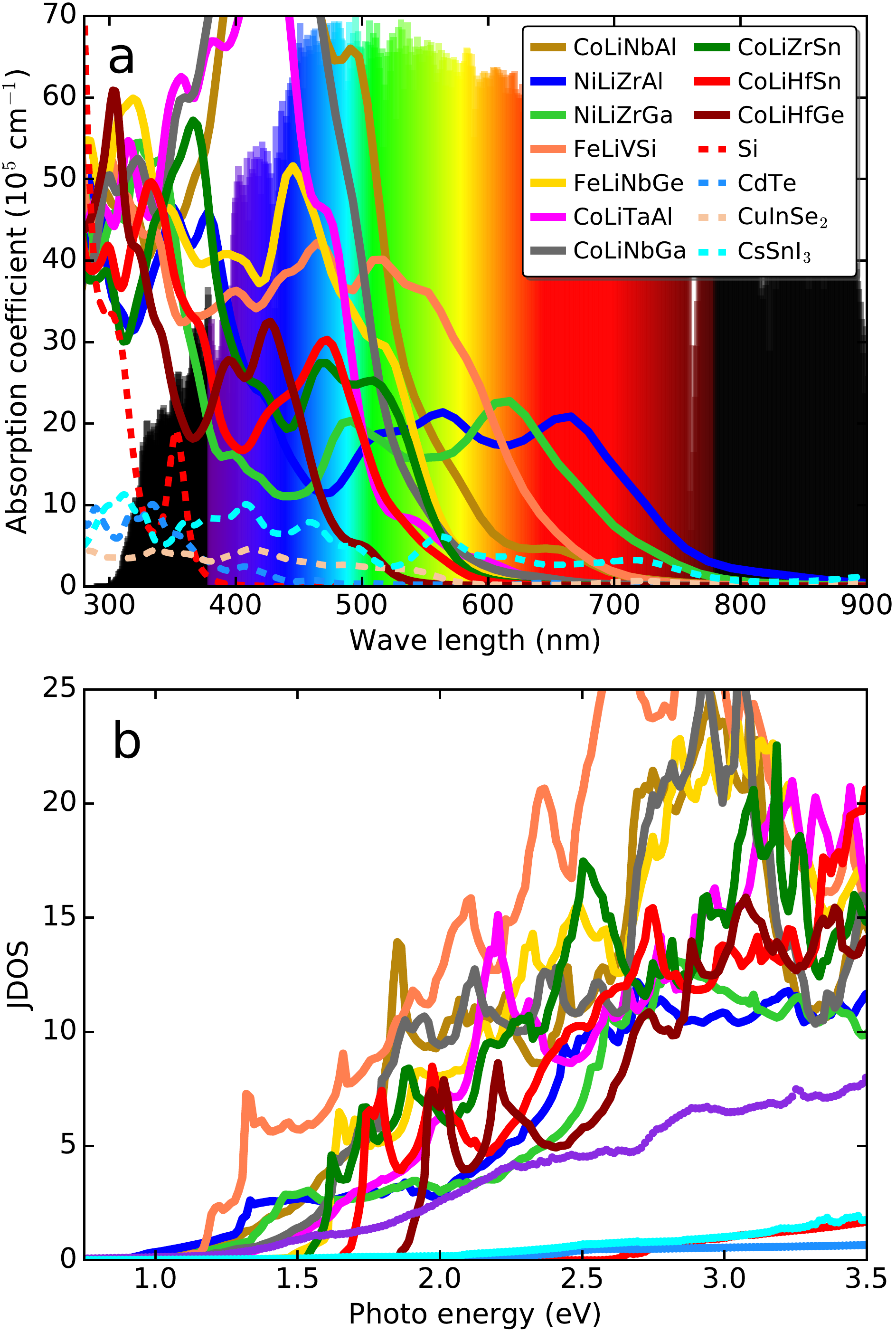}
\caption{\textbf{Optical properties of some selected quaternary Heusler
compounds.} {\bf a.} Optical absorption coefficients.
\textbf{b.} joint density of states (JDOS) of the selected quaternary
Heusler compounds, compared with conventional PV materials Si, CdTe,
CuInSe$_2$, and CsSnI$_3$.}
\label{FIG:SPECT}
\end{figure}

\section{Potential photovoltaic applications}
Photovoltaic (PV) materials, which can directly convert sunlight into electric
power, are one of the most important and promising renewable energy material
platforms. The amount of sunlight absorbed directly limits the power conversion
efficiency (PCE) of a solar cell. Since most solar radiation on the surface of the
Earth is contained within 2\,eV, high sunlight absorption in semiconducting
PV material requires its band gap to be smaller than 2\,eV. However, a high
open-circle voltage ($V_{\rm O}$) demands a large band gap. For a single $p$-$n$
junction solar cell, the band gap of the PV material must be 1.2--2.0\,eV to
maximize PCE. In this paper we focus on the optical
absorption properties and dielectric constants of the QH compounds with band
gaps between 1.0 and 2.0\,eV at the HSE06 level. The optical absorption
coefficients ($\alpha$) of such QH compounds containing earth-abundant elements
were computed using the HSE06 functional (see \textcolor{blue}{Methods} for details).

As shown in Fig.~\ref{FIG:SPECT}a, all QH compounds show very high $\alpha$ in
the visible light range, higher than other widely studied PV materials, such as
$\alpha$-silicon, CdTe, CuInSe$_2$, and CsSnI$_3$. A strong interband optical
absorption can be expected from the high joint density of state (JDOS) of the QH
compounds, in comparison to conventional PV materials (see
Fig.~\ref{FIG:SPECT}b); this follows from the imaginary part of the dielectric
function ($\epsilon_2$) being proportional to the JDOS and the momentum
transition matrix. The high JDOS of QH compounds is due to the fact that the
states at the valence band maximum (VBM) and conduction band minimum (CBM) are
mainly from the more localized $d$ states of the transition metal atom as shown
in Fig.~\ref{FIG:BAND}d, and the $d$-$d$ transition is much stronger than the
usual $s$-$p$ one. In contrast, the VBM and CBM of $\alpha$-silicon, GdTe,
CuInSe$_2$, and CsSnI$_3$ are from the more delocalized $s$ and $p$ orbitals. In
addition, the restrictions imposed by dipole selection rules are alleviated by
the broken inversion symmetry of the QH structure (F$\bar{4}3m$), which
contributes to the large $\alpha$ as well. Moreover, the structural similarity
between QH and zinc blende makes QH a suitable buffer layer for semiconductors
used in solar cells if their lattice constants match~\cite{Kieven2010}.

Other key factors required for high PCE materials are photo-generated
electron-hole pair separation, carrier mobility, and carrier diffusion length,
which are related to the exciton binding energy (E$_{\rm bin}$) and carrier
effective masses ($m^*$). In the Wannier-Mott (WM) model~\cite{wannmottexciton}, E$_{\rm
bin}$ is inversely proportional to the square of the high-frequency dielectric
constant ($\epsilon_{\infty}$) and proportional to the reduced effective mass
($\mu$) of the electron ($m_{\rm e}^*$) and hole ($m_{\rm h}^*$) pair: E$_{\rm bin}\propto
\frac{\mu}{\epsilon^2_{\infty}}$. Therefore, large $\epsilon_{\infty}$ and small
$\mu$ are favored for high PCE. Moreover, strong screening due to the large
dielectric constant significantly reduces the probability of carrier trapping by
impurities and defects, and electron-hole recombination. As listed in
Table~\ref{dielectric}, all QH semiconductors have very large values of
$\epsilon_{\infty}$, 2--4 times larger than most well studied solar cell
materials, such as Si, CdTe, CuInSe$_2$, and CsSnI$_3$ computed at the same
level of theory.

\begin{center}
\begin{table}
\caption{{\bf Electronic properties of quaternary Heuslers compounds suitable
for PV applications.} Columns are band gap (E$_{\rm g}$), the minimal effective
mass of electron ($m_e$) and hole ($m_h$), high-frequency dielectric constant
($\epsilon_{\infty}$), and lattice constant ($a$). The dielectric constants of
QH are isotropic ($\epsilon_{\infty}^{xx} = \epsilon_{\infty}^{yy} =
\epsilon_{\infty}^{zz} = \epsilon_{\infty}$) due to the symmetry.}
\vspace{0.3cm}
\begin{ruledtabular}
\begin{tabular}{p{1.5cm}p{1.3cm}p{0.8cm}p{0.8cm}p{0.8cm}p{0.8cm}p{0.8cm}p{0.8cm}}
  Compound  &E$_{\rm g}$ (eV)&  $m_{\rm e}$ &  $m_{\rm h}$  &
  $\epsilon_{\infty}$  &  $a$ (\AA)   \\
  \hline
LiNbAlCo  &  1.041  &  0.404  &  0.314  &  24.2                &   5.963        \\
LiZrAlNi  &  1.085  &  1.082  &  1.049  &  18.6                &   6.162        \\
LiZrGaNi  &  1.150  &  1.408  &  0.812  &  18.0                &   6.124        \\
LiVFeSi   &  1.191  &  0.881  &  0.929  &  24.0                &   5.615        \\
LiNbFeGe  &  1.256  &  0.528  &  0.817  &  22.9                &   5.916        \\
LiTaAlCo  &  1.460  &  1.145  &  0.349  &  21.2                &   5.944        \\
LiNbGaCo  &  1.361  &  2.694  &  0.328  &  23.0                &   5.939        \\
LiZrCoSn  &  1.486  &  2.405  &  1.407  &  18.4                &   6.300        \\
LiHfCoSn  &  1.580  &  2.084  &  1.149  &  17.6                &   6.266        \\
LiHfCoGe  &  1.755  &  1.001  &  1.067  &  16.3                &   6.016        \\
LiTiCoSi  &  1.986  &  1.251  &  1.068  &  18.0                &   5.735        \\
Si        &  1.190  &  1.315  &  0.401  &  11.2                &   5.437        \\
CdTe      &  1.090  &  0.155  &  0.792  &  7.1                 &   6.624        \\
CuInSe$_2$&  0.620  &  0.176  &  1.097  &  8.7, 8.6            & 5.873, 11.815  \\
CsSnI$_3$ &  0.840  &  0.285  &  0.179  &  7.7                 &  6.119         \\
\end{tabular}
\label{dielectric}
\end{ruledtabular}	
\end{table}
\end{center}

\section{Potential thermoelectric applications}
\begin{figure}
\centering
\includegraphics[width=1.0\linewidth]{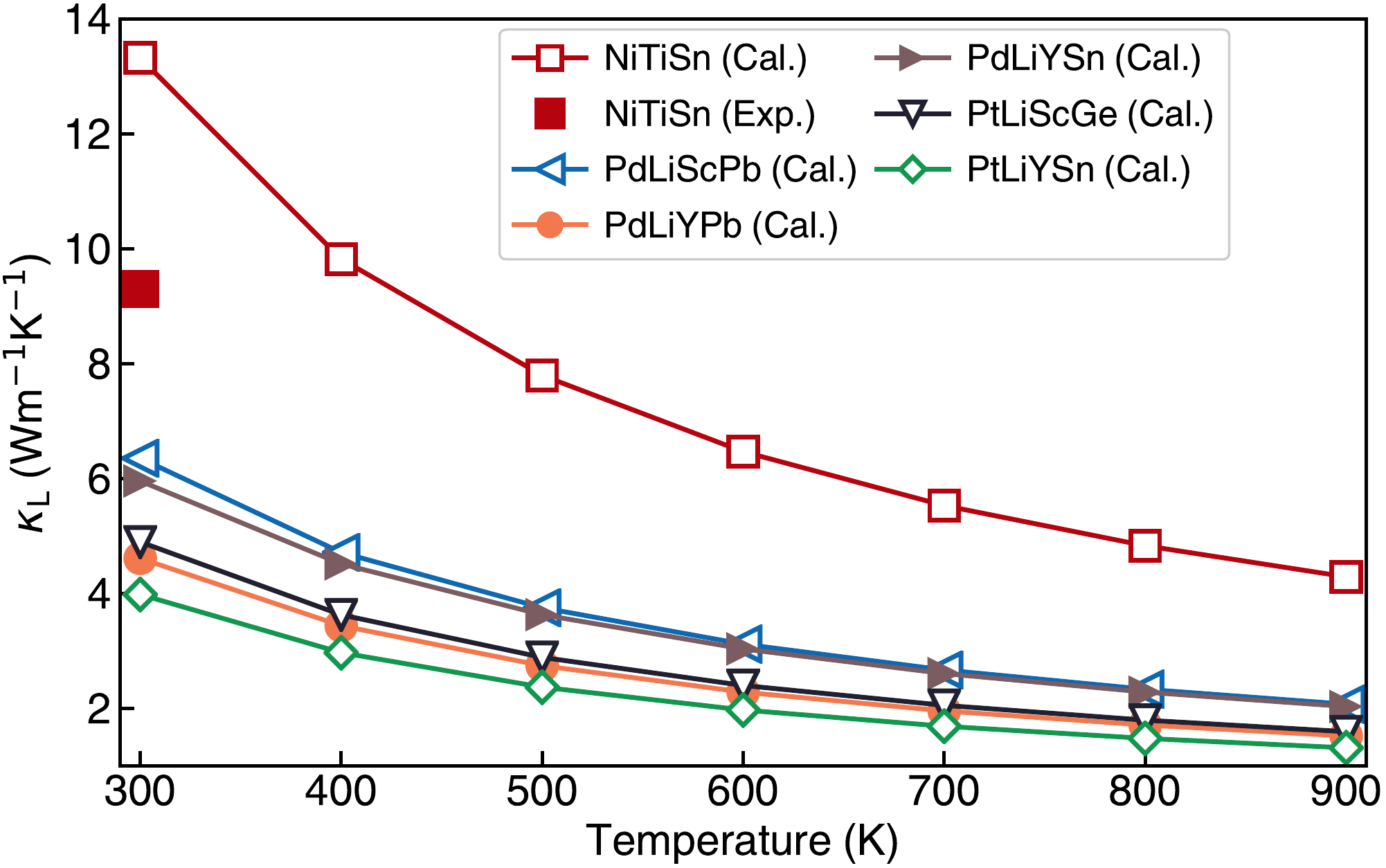}
\caption{\textbf{Lattice thermal conductivity.} The calculated (Cal.) lattice
thermal conductivity of QH compound as function of temperature. The experimental
(Exp.) and calculated (at the same level of theory) values of lattice thermal
conductivity of NiTiSn are shown for reference.}
\label{FIG:KAPPA}
\end{figure}

\begin{figure}
\centering
\includegraphics[width=1.0\linewidth]{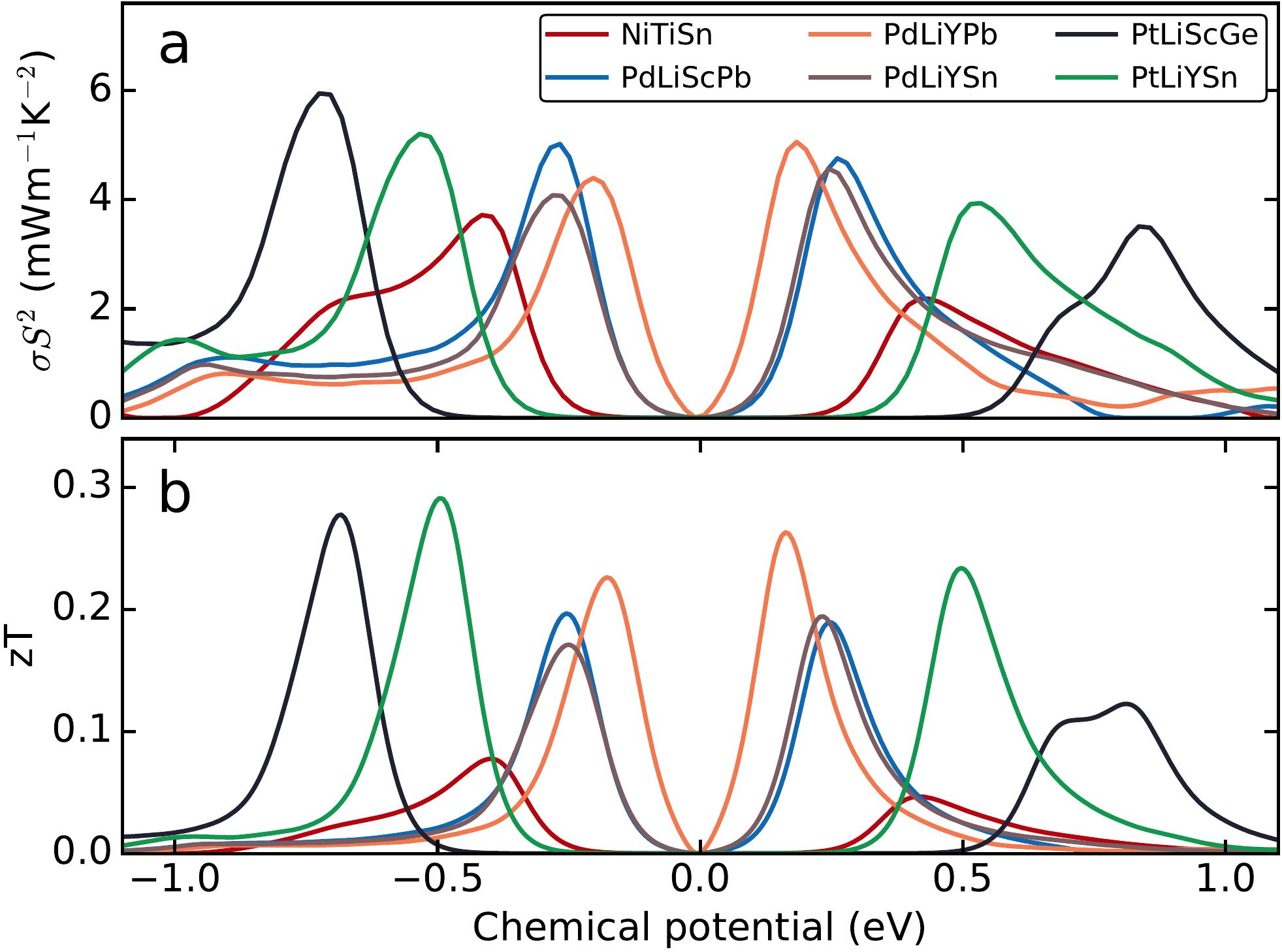}
\caption{\textbf{Calculated thermoelectric properties.} \textbf{a.} Power factor
and \textbf{b.} zT as function of chemical potential at 300\,K for QH, compared
with that calculated for NiTiSn. A fixed relaxation time $\tau=$ 15~fs was used
for all the calculations.}
\label{FIG:PF}
\end{figure}

Semiconducting Heusler (HH and FH) compounds have been widely studied as
thermoelectric materials for converting waste heat into electricity. These
Heusler compounds often possess a high power factor (PF = $\sigma{S^2}$,
$\sigma$ and $S$ are electrical conductivity and Seebeck coefficient,
respectively) stemming from the high band degeneracy ($N_{\rm v}$), high density
of states near the Fermi level, and the combination of ``flat-and-dispersive''
band structure~\cite{Bilc2015,PhysRevLett.117.046602,xie2013beneficial}.
However, the high intrinsic lattice thermal conductivity ($\kappa_{\rm
L}\geq$~10~Wm$^{-1}$K$^{-1}$ at 300\,K) significantly impedes thermoelectric
efficiency that requires a high PF and low overall thermal conductivity $\kappa
= \kappa_{\rm L} + \kappa_{\rm e}$, where $\kappa_{\rm e}$ is the electron
contribution. Therefore, semiconducting Heusler compounds with a high PF
and low intrinsic $\kappa_{\rm L}$ are very promising candidate materials for thermoelectric
applications~\cite{PhysRevLett.117.046602,PhysRevX.4.011019}. Here, we calculate
$\kappa_{\rm L}$ of the newly discovered QH semiconductors with HSE06 band gaps
E$_{\rm g}$ less than 1.2\,eV by solving the linearized Boltzmann equation based on
anharmonic third-order phonons with the ShengBTE package~\cite{ShengBTE_2014}.
As shown in Fig.~\ref{FIG:KAPPA},
the calculated $\kappa_{\rm L}$ values for some of these compounds (LiScPdPb,
LiYPdPb, LiYPdSn, LiScPtGe, and LiYPtSn) are around 4--7 Wm$^{-1}$K$^{-1}$ at
300~K, which are about two to four times lower than the widely studied HH
thermoelectric materials, such as NiTiSn, computed at the same level of theory.
We note that our calculated value of $\kappa_{\rm L}$ for NiTiSn agrees 
well with a previous calculation of 16.8 Wm$^{-1}$K$^{-1}$ at 300
K~\cite{PhysRevX.4.011019}. The computed $\kappa_{\rm L}$ values might be
overestimated (e.g., NiTiSn: 13.5 and 9.3 Wm$^{-1}$K$^{-1}$ for calculation and
experiment at 300 K, respectively) since our calculations neither include higher
order force constants nor consider effects of defects and grain boundaries which
can further scatter heat carrying phonons. We also compute the PF by solving the
electronic Boltzmann equation with the constant relaxation time approximation, as
implemented in the Boltztrap code~\cite{BoltzTrap} (see \textcolor{blue}{Methods} for computational
details). Similar to the well-studied half Heusler NiTiSn, all the QH compounds
studied in this paper have reasonably high PF for both electron and hole
transport assuming the same electron relaxation time ($\tau=$ 15~fs) as NiTiSn.
The high power factor is due to (a) conduction band minimum (CBM) and valence
band maximum (VBM) always being at low symmetry points of the Brillouin zone,
resulting in a high valley-degeneracy, and (b) high density of states near the
Fermi level at the top of valence band and bottom of conduction band mainly from
the $d$ orbitals of a transition metal (Fig.~\ref{FIG:BAND}). Based on the same
relaxation time ($\tau=$ 15~fs), we also estimate the $zT$ of these QH compounds
using the calculated $\kappa_{\rm{L}}$ values and the electronic thermal
conductivity ($\kappa_{\rm e}$). The estimated
$zT=\frac{S^2\sigma}{\kappa_{\rm e}+\kappa_{\rm L}}$ values for these QH
semiconductors at 300\,K are 2--4 times higher than NiTiSn, as shown in
Fig.~\ref{FIG:PF}. Interestingly, from Fig.~\ref{FIG:PF} we see that these QH
can possibly be used as either $n$-type or $p$-type thermoelectric materials with
appropriate doping. Similar to other Heusler compounds, the $\kappa_{\rm L}$ can
be further reduced by grain boundaries, alloying, nano-scale second-phase
particles, etc.

\section{Conclusions}
By employing high throughput DFT screening combined with phase stability
analysis within the Open Quantum Materials Database, we discover 99 new, stable,
and ordered quaternary Heusler semiconductors. Our discovery of this large set
of new stable compounds is based on the design principle of filling the voids of
the half Heusler with another atom to obtain 18 valence electrons, which is
itself a conceptual extension of creating a half Heusler from the zinc blende
structure. The inserted atom that can fill the void in the half Heusler
structure should possess small size and high electropositivity. This is in
contrast to filling the void in the zinc blende structure to generate half
Heusler compounds, where late transition metals are favorable as their nearly
full $d$ states can be filled easily. The electronic structure of the quaternary
Heusler compounds are very similar to that of half Heusler compounds, where the
valence and conduction bands are from the hybridization between $X$-$d$ and
$Z$-$p$ orbitals, and the conditions of opening a band gap is the same: 18
valence electrons per formula unit are required to fully fill 9 valence
bands.

Our initial studies on the properties of these QH compounds indicate that many of these
semiconducting intermetallics have very high optical absorption in the visible
light range and giant dielectric screening, which can significantly promote
photogenerated electron-hole pair generation and separation, and therefore are
attractive photovoltaic materials. Several QH semiconductors with smaller band
gaps are found to possess the rare combination of simultaneous high power factor
and low thermal conductivity, indicating promise for thermoelectric
applications. Our discovery opens up possibilities for synthesizing and
designing novel materials for various applications in an accelerated
fashion.

\section{Methods}
\label{sec:methods}
All the HT calculations are based on DFT as implemented in Vienna Ab initio
Simulation Package (VASP)\cite{vasp1,vasp2}. The Perdew-Burke-Ernzerhof (PBE)
exchange-correlation functional~\cite{PBE} and plane wave basis set were used.
The qmpy~\cite{OQMD} framework was used for HT-DFT screening and phases from the
Open Quantum Material Database (OQMD)~\cite{OQMD} were used for convex hull
construction. The lowest energy structure of $X$Li$YZ$ was confirmed by using
minima hopping crystal structure prediction method~\cite{mhm1,mhm2} and
particle-swarm optimization implemented in the CALYPSO
package~\cite{PhysRevB.82.094116,wang2012calypso} using a 16-atoms unit cell.
Lattice dynamic stability was investigated by performing frozen phonon
calculations using the phonopy package~\cite{phonopy2}. The site
occupancy disorder in QH was investigated by comparing the energy of the ordered
QH with the disordered QH simulated using special quasirandom structures (SQS).

The band gap, optical properties, and dielectric constants were computed by
means of the screened hybrid functional HSE06~\cite{heyd2003hybrid}.
The high-frequency dielectric constant is calculated by using the perturbation
expansion after discretization (PEAD) method~\cite{PhysRevB.63.155107,PhysRevLett.89.117602}.
The absorption coefficient was calculated by using following formula
\[\alpha{(\omega)} = \frac{2\omega{\sqrt{\frac{\sqrt{\epsilon_1^2(\omega)+\epsilon_2^2(\omega)} - \epsilon_1(\omega)}{2}}}}{\rm c}\,,
\]
where $\epsilon_1$ and $\epsilon_2$ are the real and imaginary parts of dielectric function calculated at the HSE06 level, respectively,
$\omega$ is the photon frequency, and c is the speed of light.

Electron transport properties were calculated by using Boltztrap code within relaxation
time approximation~\cite{BoltzTrap} with a 41$\times$41$\times$41 \textbf{k}-mesh.
Electron thermal conductivity ($\kappa_{\rm e}$) is calculated from $\sigma$, using the Wiedemann-Franz law ($\kappa_{\rm e}={\rm L_0}\sigma$T) with Lorenz number $\rm L_0=2.44\times$10$^{-8}~$W$\Omega$K$^{-2}$~\cite{Franz1853}. 
Lattice thermal conductivities were calculated with the third-order phonon approach~\cite{PhysRevB.86.174307}, with the linearized Boltzmann equation solved
using the ShengBTE package~\cite{ShengBTE_2014}.
The second and third-order interatomic force constants were calculated with a real-space
supercell approach~\cite{phonopy2,PhysRevB.86.174307}.
The $\kappa_{\rm L}$ calculation was performed with a 24$\times$24$\times$24 \textbf{q}-mesh, see converge test in Supplementary Figs.~\textcolor{blue}{S4}.

\bibliographystyle{aipnum4-1}   
\bibliography{Ref}

\section{Acknowledgments}
J.H. (HT DFT, phonon, crystal structure search, optical and thermoelectric properties)
and C.W. (conceived and designed the project) acknowledge support via ONR STTR N00014-13-P-1056.
S.S.N. (optical and thermoelectric properties) was supported by US Department of Energy,
Office of Science, Basic Energy Sciences, under grant DEFG02-07ER46433.
V.H. (HT DFT) acknowledges support from the Department of Energy, under grant DE-SC0015106.
M.A. (Minima Hopping) acknowledges support from the Novartis
Universit\"{a}t Basel Excellence Scholarship for Life Sciences,
the Swiss National Science Foundation (P300P2-158407, P300P2-174475).
This research was supported by resources from the National Energy Research Scientific Computing Center,
a DOE Office of Science User Facility supported by the Office of Science of the U.S. Department of Energy under
Contract No. DE-AC02-05CH11231, the Extreme Science and Engineering Discovery Environment (XSEDE) (which is supported by National Science Foundation Grant No. OCI-1053575), and the Quest high-performance computing facility at Northwestern University.


\section{Author contributions}
The research was conceived and designed by J.H. and C.W. High throughput DFT
calculations were carried out by J.H. and V.H. Photovoltaic, dielectric
constant, and thermoelectric properties calculations were conducted by J.H. and
S.S.N. Minima Hopping crystal structure searches were set up by M.A. Analysis
of the data was performed by J.H. and S.S.N. 
Lattice thermal conductivity calculations were carried out by J.H., S.S.N., and M.A.
All authors discussed the results contributed to writing the manuscript.

\vspace{0.5cm}
\section{Additional information}
Supplementary information is available in the online version of the paper.

\section{Competing financial interests}
The authors declare no competing financial interests.

\end{document}